\documentclass[]{jfm}

\usepackage{graphicx}
\usepackage{newtxtext}
\usepackage{newtxmath}
\usepackage{natbib}
\usepackage{hyperref}
\hypersetup{
    colorlinks = true,
    urlcolor   = blue,
    citecolor  = black,
}

\newcommand{\RomanNumeralCaps}[1]
\linenumbers

\newcommand{\kk}{\boldsymbol{k}}
\newcommand{\eex}{\boldsymbol{e}_x}
\newcommand{\eey}{\boldsymbol{e}_y}
\newcommand{\eez}{\boldsymbol{e}_z}
\newcommand{\diff}{\mathrm{d}}
\newcommand{\bOmega}{\boldsymbol{\Omega}}
\newcommand{\bnu}{\boldsymbol{\nu}}
\newcommand{\bmu}{\boldsymbol{\mu}}
\newcommand{\balpha}{\boldsymbol{\alpha}}
\newcommand{\bbeta}{\boldsymbol{\beta}}
\newcommand{\bchi}{\boldsymbol{\chi}}
\newcommand{\bd}{\boldsymbol{d}}

\newcommand{\vvk}{\hat{\boldsymbol{u}}_{\kk}}
\newcommand{\vvp}{\hat{u}_{p\kk}}
\newcommand{\vvt}{\hat{u}_{t\kk}}
\newcommand{\vvs}{\hat{\boldsymbol{u}}_{s\kk}}
\newcommand{\bbk}{\hat{b}_{\kk}}

\newcommand{\thk}{\theta_{\kk}}
\newcommand{\phk}{\varphi_{\kk}}
\newcommand{\ok}{\omega_{\kk}}
\newcommand{\tho}{\theta_{1}}
\newcommand{\pho}{\varphi_{1}}
\newcommand{\oo}{\omega_{1}}
\newcommand{\tht}{\theta_{2}}
\newcommand{\pht}{\varphi_{2}}
\newcommand{\ot}{\omega_{2}}

\newcommand{\eek}{\boldsymbol{e}_{\boldsymbol{k}}}

\newcommand{\eep}{\boldsymbol{e}_{p\kk}}
\newcommand{\eepo}{\boldsymbol{e}_{p1}}
\newcommand{\eept}{\boldsymbol{e}_{p2}}
\newcommand{\eet}{\boldsymbol{e}_{t\kk}}

\newcommand{\ak}{a_{\kk}}
\newcommand{\acmk}{a_{-\kk}^*}
\newcommand{\aco}{a_1^*}
\newcommand{\act}{a_2^*}

\newcommand{\Votk}{V_{12}^{\kk}}
\newcommand{\Vkto}{V_{\kk2}^{1}}
\newcommand{\Vkot}{V_{\kk1}^{2}}

\newcommand{\R}{\mathcal{R}}
\newcommand{\Q}{\mathcal{Q}}
\newcommand{\D}{\mathcal{D}}
\newcommand{\I}{\mathcal{I}}
\newcommand{\J}{\mathcal{J}}
\newcommand{\T}{\mathcal{T}}

\newcommand{\Rotk}{\R_{12}^{\kk}}
\newcommand{\Rkto}{\R_{\kk2}^{1}}
\newcommand{\Rkot}{\R_{\kk1}^{2}}
\newcommand{\Qotk}{\Q_{12}^{\kk}}
\newcommand{\Qkto}{\Q_{\kk2}^{1}}
\newcommand{\Qkot}{\Q_{\kk1}^{2}}

\newcommand{\Rotkh}{\R_{12}^{\kk(h)}}
\newcommand{\Rktoh}{\R_{\kk2}^{1(h)}}
\newcommand{\Rkoth}{\R_{\kk1}^{2(h)}}

\newcommand{\Totk}{\T_{12}^{\kk}}

\newcommand{\ra}{r_{\alpha}}
\newcommand{\rb}{r_{\beta}}
\newcommand{\rc}{r_{\gamma}}
\newcommand{\rd}{r_{\delta}}

\title{On the kinetics of internal gravity waves beyond the hydrostatic regime}

\author{Vincent Labarre\aff{1}
	\corresp{\email{vincent.labarre@oca.eu}},
	Nicolas Lanchon\aff{2} \corresp{\email{nicolas.lanchon@universite-paris-saclay.fr}},
	Pierre-Philippe Cortet\aff{2} \corresp{\email{pierre-philippe.cortet@universite-paris-saclay.fr}},
	Giorgio Krstulovic\aff{1} \corresp{\email{giorgio.krstulovic@oca.eu}},
	\and Sergey Nazarenko\aff{3} \corresp{\email{sergey.nazarenko@unice.fr}}}

\affiliation{\aff{1} Universit\'{e} C\^{o}te d'Azur, Observatoire de la C\^{o}te d'Azur, CNRS, Laboratoire Lagrange, Nice, France
	\aff{2} Universit\'{e} Paris-Saclay, CNRS, FAST, 91405 Orsay, France
	\aff{3} Universit\'{e} C\^{o}te d'Azur, CNRS, Institut de Physique de Nice - INPHYNI, Nice, France}

\begin{document}
\maketitle

\begin{abstract}
	We present a new derivation of the kinetic equation for weak, non-hydrostatic internal gravity wave turbulence. The equation is equivalent to the one obtained by \citet{caillol_kinetic_2000}, but it takes a canonical form. We show that it conserves the energy without involving the resonance condition in frequency, and look for the isotropic part of the steady, scale invariant solutions. We provide a parametrization of the resonant manifold of non-hydrostatic internal gravity wave triadic interactions. This allows us to simplify the collision integral, and to evaluate the transfer coefficients of all triadic interactions. In the hydrostatic limit, our equation is equivalent to the Hamiltonian description of \citet{lvov_hamiltonian_2001}.  
\end{abstract}


\section{Introduction}
\label{section1}

Internal gravity waves (IGW) propagate in stably stratified fluids. They are the consequence of a restoring buoyancy force that makes fluid particles oscillate around their floatability level. IGW can be found in various environments, including the atmosphere, oceans, lakes, rivers, and industrial flows \citep{lelong_riley_internal_1991,staquet_sommeria_internal_2002}. They can be excited by wind shear, buoyancy forcing due to heating, topography, or tides \citep{vallis_atmospheric_2017}. They are important because they transport mass, momentum, and energy, which can significantly impact the large-scale features of stratified flows. The mutual nonlinear interactions between IGW and with other flow components can lead to the transfer of energy between different scales. This results in the generation of smaller-scale waves and vortices. \\

IGW have been studied from a variety of perspectives, including observations \citep{mackinnon_parametric_2013}, experiments \citep{rodda_experimental_2022, lanchon_internal_2023}, numerical simulations \citep{pan_numerical_2020,lam_energy_2021}, and theoretical models \citep{caillol_kinetic_2000, lvov_hamiltonian_2001, dematteis_downscale_2021}. Under the Boussinesq approximation, three key non-dimensional numbers drive stratified flows. Namely, the Froude number $Fr = U/(NL)$, the Reynolds number $\Rey=UL/\nu $, and the Prandtl number $\Pran=\nu/\kappa$, where $U$ is the typical velocity of the flow, $L$ the size of the domain, $N$ the Brunt-Väisälä frequency, $\nu$ is the viscosity, and $\kappa$ is the scalar diffusivity. When the flow is composed of waves only, being strongly stratified (i.e., $Fr \ll 1$), with the viscosity and diffusivity being small (i.e., $\Rey, \Pran \Rey \gg 1$) but such that the dynamics of energetic modes remain weakly nonlinear, a regime described by the Weak Wave Turbulence (WWT) theory is foreseen \citep{hasselmann_feynman_1966,zakharov_kolmogorov_1992,nazarenko_wave_2011,galtier_physics_2022}. In such a state, the energy is concentrated on the linear dispersion relation for a continuum range of scales, as observed in the numerical simulations of \citet{reun_parametric_2018}. It is characterised by interactions over wave triads satisfying resonance conditions in wave vectors and frequencies. Note that in this theoretical description, we do not consider a shear flow and vortical modes (with vorticity parallel to the vertical axis), so the typical velocity $U$ corresponds to velocity fluctuations of weakly nonlinear waves and not to a mean flow velocity. \\

The WWT theory for geophysical flows was developed notably in the 1960s by \citet{hasselmann_feynman_1966}, who derived a general kinetic equation using a Lagrangian formalism for several wave systems \citep{hasselmann_nonlinear_1967}, including IGW. \citet{muller_dynamics_1975, olbers_nonlinear_1976} extended the work of Hasselmann to write the first kinetic equation for internal waves, i.e. with rotation and stratification. Since then, the kinetic equation for internal waves, with or without rotation and in or outside the hydrostatic limit, has been re-derived using various approaches: Clebsch variables \citep{pelinovsky_raevsky_weak_1977, voronovich_hamiltonian_1979}; a decomposition between vertical velocity, potential part of the horizontal velocity, and vertical vorticity \citep{caillol_kinetic_2000,caillol_erratum_2001}; and isopycnal coordinates \citep{lvov_hamiltonian_2001,lvov_hamiltonian_2004,medvedev_turbulence_2007}. We refer the reader to \citet{lvov_resonant_2012} for a review of earlier IGW kinetic equations. Three classes of triadic interactions corresponding to non-local transfers were identified \citep{mccomas_bretherton_resonant_1977}. Induced diffusion (ID) occurs when one low-frequency wave interacts with two approximately identical waves of much larger wave number and frequency. Elastic scattering (ES) occurs when two waves which are nearly vertical reflections of each other interact with a third wave which has a low-frequency and almost twice the vertical wave number of the other two waves. Finally, the parametric subharmonic instability mechanism (PSI) is an instability wherein a low wave number wave decays into two high wave vector waves of half its frequency. Recently, it has been found that local interactions, i.e. involving waves with similar wave-vector amplitudes are also very important in the energy transfers \citep{dematteis_downscale_2021, wu_energy_2023}, particularly those which lie on the same vertical plane. \\

WWT gives predictions, among other things, for the wave energy spectrum $e_{\kk}$. Usually, the theory is formulated in terms of the wave-action spectrum $n_{\kk} = e_{\kk} / \ok$, $\ok$ being the wave frequency. Physically, $n_{\kk}$ is interpreted as the number of waves with wave-vector $\kk$. The theory gives the kinetic equation $\dot{n}_{\kk} = St_{\kk}$, $St_{\kk}$ being the collision integral, describing the dynamics of the wave action spectrum on long time scale due to wave-wave interactions. In particular, axisymmetric, bihomogeneous, steady-state wave action spectra $n_{\kk} \propto k_h^{\nu_h} |k_z|^{\nu_z}$ were previously obtained as solutions of the kinetic equation in the hydrostatic limit. They correspond to the zeros of the collision integral, including the thermal equilibrium spectrum (often called the Rayleigh-Jeans spectrum) or spectra associated with a non-zero energy flux (called Kolmogorov-Zakharov (KZ) spectra). Using energy conservation, one KZ spectrum was found analytically $n_{\kk} \propto k_h^{-7/2} |k_z|^{-1/2}$. It is known as the Pelinovsky-Raevsky (PR) spectrum \citep{pelinovsky_raevsky_weak_1977, caillol_kinetic_2000,lvov_hamiltonian_2001}. However, as we will explain in subsection \ref{subsection3d}, it has been found that this spectrum is not a mathematically valid solution \citep{caillol_kinetic_2000,lvov_oceanic_2010}, and is thus physically irrelevant. Using numerical integration, $n_{\kk}\propto k_h^{-3.69}$ actually turned out to be the only zero of the collision integral \citep{lvov_oceanic_2010,dematteis_downscale_2021}. More recently, \citet{lanchon_energy_2023} derived a stationary solution ($n_{\kk}\propto k_h^{-3} |k_z|^{-1}$) of a simplified kinetic equation describing the small scales of the nonlocal internal wave turbulence problem. This derivation is based on the assumption that energy transfers are driven only by ID triads, which lead to a scale separation in the kinetic equation. \\

As said earlier, there are already four ways of deriving the kinetic equation of IGW. The Lagrangian approach \citep{hasselmann_nonlinear_1967,olbers_nonlinear_1976} has the advantage of being non-hydrostatic and takes into account rotation. Yet, the incompressibility has to be treated perturbatively and some computations could be simplified by exploiting Hermitian symmetries. For more, even if it is not the purpose of the present study, the Lagrangian formalism is not adapted if we want to include vertical vorticity (or geostrophic modes when rotation is added) \citep{caillol_kinetic_2000}. The approaches using Clebsch variables \citep{pelinovsky_raevsky_weak_1977, voronovich_hamiltonian_1979} have the advantage of being Hamiltonian. However, the decomposition does not apply to fields with vertical vorticity, the physical meaning of the conjugate variables is less straightforward, and the references are not easily available. The isopycnal coordinates \citep{lvov_hamiltonian_2001,lvov_energy_2004,medvedev_turbulence_2007} have the advantages of being a canonical Hamiltonian description of the flow and to be able to take into account vertical vorticity (or geostrophic modes), but only in the hydrostatic limit. \citet{milder_hamiltonian_1982} gave a Hamiltonian description of the flow holding outside the hydrostatic limit, but it is not canonical and the kinetic equation of IGW was not derived. The decomposition used by \citet{caillol_kinetic_2000} is the closest in spirit to the present study. It has the advantage of being derived in the very common Eulerian coordinates system, to allow the description of vortical modes, and to be valid outside the hydrostatic limit. Yet, the resulting kinetic equation was not shown to have a canonical structure. We show here that it turns out to be the case (after accounting for misprints which were corrected in Erratum \cite{caillol_erratum_2001}). Here we use the poloidal-toroidal decomposition \citep{godeferd_toroidal_2010}, also known as the Craya-Herring decomposition \citep{craya_contribution_1957,herring_approach_1974}. As explained later, it is particularly adapted to the derivation of the kinetic equation of IGW. Namely, it uses the common Eulerian coordinates system, offers a complete basis of the components of stratified flows, and takes into account incompressibility from the beginning. We also simplify the notations when compared to the Lagrangian formalism by exploiting the Hermitian symmetry satisfied by the velocity and buoyancy fields. These advantages allow us to recast the kinetic equation of IGW into a canonical form, that is more amenable to analytical and numerical treatments. \\

The remainder of the paper is as follows. In the next section \ref{section2}, we present the poloidal-toroidal decomposition, which is very convenient for studying IGW and for writing the equations of motion in the canonical variables. Section \ref{section3} is devoted to the kinetic description of weak IGW. The kinetic equation is derived in subsection \ref{subsection3a} using standard assumptions of WWT. In subsection \ref{subsection3b}, we look for steady, scale invariant solutions to the kinetic equation. We show in subsection \ref{subsection3c} that, when evaluated on the resonant manifold, the interaction coefficients are symmetric with respect to permutation of the wave-vectors. It allows us to write the canonical form of the kinetic equation. We also parametrize the resonant manifold, which allows us to give a simplified version of the collisional integral for axisymmetric spectra, and to study the transfer coefficients of triadic interactions. We study the hydrostatic limit in subsection \ref{subsection3d}. In that limit, our kinetic equation is equivalent to many previous ones in that case. We show that the PR spectrum \citep{pelinovsky_raevsky_weak_1977} can be obtained without using the frequency resonance condition, which, up to our knowledge, was not remarked before. We conclude in section \ref{section4}. Technical details about the derivation of the kinetic equation are presented in Appendix \ref{appendix1}.

\section{Equations of motion}
\label{section2}

We start from the three-dimensional Boussinesq equations:
\begin{align}
	\label{eq:Continuity}
	\nabla \cdot \boldsymbol{u}&= 0, \\
	\label{eq:Impulsion}
	\p_t\boldsymbol{u}+ \boldsymbol{u}\cdot \nabla \boldsymbol{u}&= - \nabla p + b ~ \eez, \\
	\label{eq:Buoyancy}
	\p_t{b} + \boldsymbol{u}\cdot \nabla b &= -N^2 u_z,
\end{align}
where $(x,y,z)$ denote the three spatial coordinates in the Cartesian frame $(O, \eex, \eey, \eez)$, $\eez$ is the unitary vector along the stratification axis, $\boldsymbol{u}=(u_x, u_y, u_z)$ the velocity, $b$ the buoyancy, $p$ the total kinematic pressure, and $N$ the Brunt-Väisälä
(or buoyancy) frequency. The buoyancy is defined as $b = -g \rho' / \rho_0$, where $g$ is the acceleration due to gravity, $\rho_0$ is the average density of the fluid at $z=0$, and $\rho'$ is the density perturbation with respect to the average linear density profile $\bar{\rho}(z) = \rho_0 + \frac{\diff \bar{\rho}}{\diff z}z$. It follows that $N = \sqrt{-\frac{g}{\rho_0}\frac{\diff \bar{\rho}}{\diff z}}$. Equations (\ref{eq:Continuity}-\ref{eq:Buoyancy})
conserve the total energy $E$ and the potential vorticity $\Pi$ is a Lagrangian invariant \citep{bartello_geostrophic_1995}, with
\begin{equation}
	\label{eq:Invariants}
	E \equiv \frac{1}{L^3} ~ \int ~ \left[ \frac{\boldsymbol{u}^2}{2} + \frac{b^2}{2N^2} \right] ~ \diff x \diff y \diff z, ~~~~ \Pi = \bOmega \cdot \left( N^2 \eez + \nabla b \right),
\end{equation}
and $\bOmega = \nabla \times \boldsymbol{u}$ being the vorticity. \\

\begin{figure}
	\centerline{\includegraphics[width=0.4\textwidth]{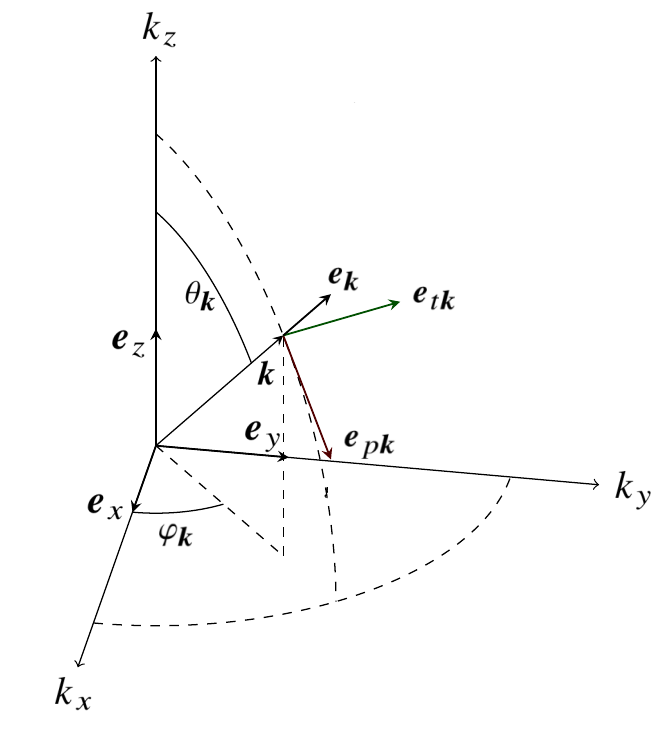}} 
	\caption{Illustration of the poloidal-toroidal basis $(\eek, \eep, \eet)$ defined by
		equations (\ref{eq:poloidal-toroidal}). $\thk$ is the polar angle (i.e., between $\eez$ and $\eek$), and $\phk$ is the azimuthal angle (i.e., between the horizontal projection of $\eex$ and $\kk$).
		\label{figure1}}
\end{figure}

We consider a triply-periodic domain with spatial periods $L_x = L_y = L_z = L$. The Fourier transform of the velocity field $\vvk = (\hat{u}_{x \kk}, \hat{u}_{y \kk}, \hat{u}_{z \kk})$
can be written using the poloidal-toroidal-shear decomposition \citep{craya_contribution_1957,herring_approach_1974,godeferd_toroidal_2010}, which is now common in the study of stratified flows. Namely, we have
\begin{equation}
	\vvk = \begin{cases} \vvp ~ \eep + \vvt ~ \eet ~~~~ \text{if} ~ k_h \neq 0 \\
		\vvs = \hat{u}_{x\kk} ~ \eex + \hat{u}_{y\kk} ~ \eey ~~~~ \text{if} ~ k_h = 0 
	\end{cases},
\end{equation}
where
\begin{equation}
	\label{eq:poloidal-toroidal}
	\eek = \frac{\kk}{k}, ~~~~
	\eep = \frac{\kk \times (\kk \times \eez)}{|\kk \times (\kk \times \eez)|}, ~~~~
	\eet = \frac{\eez \times \kk}{|\eez \times \kk|}, 
\end{equation}
$\vvp$ is the poloidal component, $\vvt$ the toroidal component, $\vvs$ the shear
modes component, $\kk=(k_x, k_y, k_z)$ denotes the wave vector, $k =|\kk| = \sqrt{k_x^2 + k_y^2 + k_z^2}$ its modulus, and $k_h = \sqrt{k_x^2 + k_y^2}$. The basis $(\eek, \eep, \eet)$ is shown on Figure~\ref{figure1}. \\

In Fourier space, the equations of motion can be written as follows,
\begin{equation}
	\label{eq:SpectralPoloidalToroidalBuoyancyShear}
	\begin{cases}
		\dot{\hat{u}}_{p\kk} = - (\widehat{\boldsymbol{u}\cdot \nabla \boldsymbol{u}})_{\kk} \cdot \eep - \bbk \sin \thk \\
		\dot{\hat{u}}_{t\kk} = - (\widehat{\boldsymbol{u}\cdot \nabla \boldsymbol{u}})_{\kk} \cdot \eet \\
		\dot{\hat{b}}_{\kk} = - (\widehat{\boldsymbol{u}\cdot \nabla b})_{\kk} + N^2 \vvp \sin \thk
	\end{cases}
	~~~~ \text{if} ~ k_h \neq 0 ~~~~ \text{and} ~~~~
	\begin{cases}
		\dot{\hat{\boldsymbol{u}}}_{s\kk} = - (\widehat{\boldsymbol{u} \cdot \nabla \boldsymbol{u}_h})_{\kk} \\
		\dot{\hat{b}}_{\kk} = - (\widehat{\boldsymbol{u} \cdot \nabla b})_{\kk} 
	\end{cases}
	~~~~ \text{if} ~ k_h = 0
\end{equation}
where $\boldsymbol{u}_h = (u_x,u_y,0)$ is the horizontal component of $\boldsymbol{u}$. In the linear regime, the poloidal velocity and the buoyancy are coupled, while the toroidal velocity and shear modes are decoupled and not evolving. The coupling between poloidal velocity and buoyancy is responsible for the propagation of internal gravity waves with frequency 
\begin{equation}
	\pm \ok \equiv \pm N \sin \thk = \pm N \frac{k_h}{k}	
\end{equation}
where $\thk$ is the
polar angle (Figure~\ref{figure1}). Equations
(\ref{eq:SpectralPoloidalToroidalBuoyancyShear})
show that both shear modes and toroidal components have zero frequency in the linear regime, and hence they are not waves. More precisely, in the context of stratified turbulence, $\vvp$ is the kinetic part of linear waves (horizontal and vertical oscillations) while $\vvt$ corresponds to vertical vortices. It makes the poloidal-toroidal-shear decomposition particularly suitable for studying internal gravity waves and, more generally, flows with statistical axisymmetry \citep{godeferd_toroidal_2010,yokoyama_energy_2019,maffioli_signature_2020}. When compared to \citet{caillol_kinetic_2000}, the poloidal velocity encompass both the potential part of the horizontal velocity and the vertical velocity, which simplifies the computations. \\

In this study, we consider a flow composed of waves only, such that there are no toroidal and shear components. Neglecting the toroidal component is equivalent to neglecting vertical vorticity \citep{caillol_kinetic_2000}, or neglecting the ``vortex'' part in the ``wave-vortex'' decomposition in isopycnal coordinates \citep{lvov_hamiltonian_2001}. Ignoring shear and toroidal components constitute standard assumptions in IGW turbulence theory. Since the poloidal velocity and the buoyancy are real-valued, their Fourier transforms satisfy the Hermitian symmetry. This symmetry allows us, as customary in WWT, to define the general complex wave mode $\ak = \frac{1}{\sqrt{2\ok}} \left(\vvp - i \frac{\bbk}{N}\right)$, which fully determines the wave dynamics (including positive and negative frequency branches). The poloidal and buoyancy modes then express as follows

\begin{equation}
	\vvp \equiv \sqrt{\frac{\ok}{2}} \left( \ak + \acmk \right) ~~~~ \text{and} ~~~~ \frac{\bbk}{N} \equiv i \sqrt{\frac{\ok}{2}}  \left( \ak - \acmk \right).
\end{equation}
The dynamical equation then reads
\begin{equation}
	\label{eq:WaveAction}
	\dot{a}_{\kk} = - i \ok \ak - i \sum\limits_{1,2} ~ \Votk ~ \delta_{12}^{\kk} ~ a_1 a_2 - 2 i \sum\limits_{1,2} ~ \Votk ~ \delta_{\kk2}^1 ~ a_1 \act - i \sum\limits_{1,2} ~ \Votk ~ \delta_{12\kk} ~ \aco \act
\end{equation}
with the interaction coefficients
\begin{equation}
	\label{eq:InteractionCoefficients}
	V_{\kk_1 \kk_2}^{\kk} \equiv \Votk = \sqrt{\frac{\oo \ot}{32 \ok}} \left[ \left( \eepo \cdot \kk \right) \left( \eept \cdot \eep \right) + \eepo \cdot \kk_2 + \left( \eept \cdot \kk \right) \left( \eepo \cdot \eep \right) + \eept \cdot \kk_1 \right]
\end{equation}
and $\delta_{12}^{\kk}$ (respectively $\delta_{12\kk}$) being the Kronecker symbol enforcing the condition $\kk_1 + \kk_2 = \kk$ (respectively $\kk_1 + \kk_2 + \kk = 0$). Note that equation (\ref{eq:WaveAction}) is equivalent to the Boussinesq equations without toroidal velocity and shear modes. The sums represent nonlinear interaction between the wave modes. This equation has a convenient structure for deriving the wave-kinetic equation. \\

By construction, the interaction coefficients $\Votk$ are symmetric with respect to the permutation of lower indices, i.e. $\Votk = V_{21}^{\kk}$, but a priori not symmetric with respect to the permutation between a lower index and the upper index. This complication prevents us to write a canonical Hamiltonian equation for $\ak$. It is easy to show that the frequency is homogeneous, $\omega_{\mu \kk} = \mu^\alpha \ok$, with homogeneity degree $\alpha=0$. Also, the interaction coefficients are homogeneous, i.e. $V_{\mu \kk_1 \mu \kk_2}^{\mu \kk} = \mu^\beta \Votk$, with homogeneity degree $\beta=1$. Using the fact that $\eep \cdot \kk = 0$, it is easy to show that the interaction coefficients satisfy the following useful relations
\begin{equation}
	\label{eq:SymmetriesInteractionCoefficients}
	\delta_{12}^{\kk} \left( \ok \Votk - \oo \Vkto - \ot \Vkot  \right) = 0 ~~~~ \text{and} ~~~~ \delta_{12\kk} \left( \ok \Votk + \oo \Vkto + \ot \Vkot  \right) = 0.
\end{equation} 
Note that such relations between interaction coefficients are common to fluid dynamical systems with quadratic invariants (see Appendix B of \cite{remmel_new_2009} for a general proof). It allows us to prove that the wave action equation (\ref{eq:WaveAction}) conserves the energy (\ref{eq:Invariants})
\begin{equation}
	E = \sum\limits_{\kk} ~ \left[ \frac{|\vvk|^2}{2} + \frac{|\bbk|^2}{2N^2} \right] = \sum\limits_{\kk} ~ \ok |\ak|^2.
\end{equation}

\begin{figure}
	\centerline{\includegraphics[width=1\textwidth]{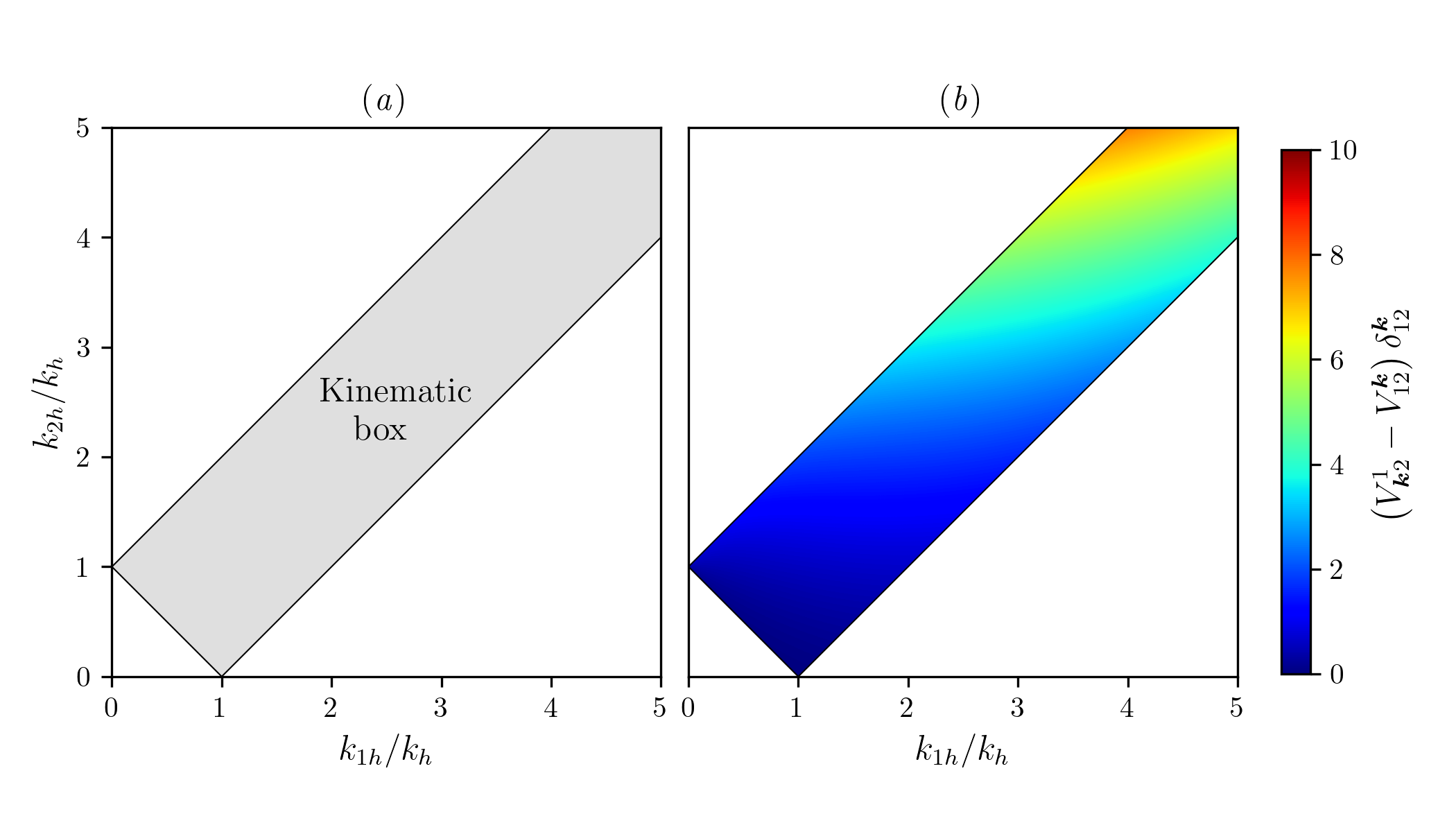}} 
	\caption{(\textit{a}) Kinematic box defined by the conditions (\ref{eq:KinematicBox}). (\textit{b}) Difference between the interaction coefficients $\Votk$ and $\Vkto$ when evaluated on the manifold $\kk = \kk_1 + \kk_2$ for $(k_h=1, k_{1h}, k_{2h}, \tho = \pi/4, \tht=\pi/6)$ and $N=1$.}
	\label{figure2}
\end{figure}

On the manifold $\kk = \kk_1 + \kk_2$ (or permutations), the interaction coefficients are only functions of $(k_h,k_{1h},k_{2h},k_{1z},k_{2z})$ or, alternatively, of $(k_h,k_{1h},k_{2h},\tho,\tht)$. Because $(\kk_h,\kk_{1h},\kk_{2h})$ form a triangle, they must satisfy the triangular inequalities
\begin{equation}
	\label{eq:KinematicBox}
	k_h \leq k_{1h} + k_{2h}, ~~~~ k_{1h} \leq k_h + k_{2h}, ~~~~ k_{2h} \leq k_h + k_{1h},
\end{equation}
meaning that $(k_{1h},k_{2h})$ must lie in the so-called ``kinematic box'' \citep{lvov_resonant_2012} shown on Figure~\ref{figure2}~(\textit{a}). We have checked numerically that the interaction coefficients are not fully symmetric with respect to permutations of indices, i.e. $\Votk ~ \delta_{12}^{\kk} \neq \Vkto ~ \delta_{12}^{\kk} \neq \Vkot ~ \delta_{12}^{\kk}$, as shown in Figure~\ref{figure2}~(\textit{b}). 

\section{Kinetic description}
\label{section3}

\subsection{Wave kinetic equation}
\label{subsection3a}

The derivation of the kinetic equation is a long, but standard exercise in WWT \citep{hasselmann_feynman_1966, zakharov_kolmogorov_1992, nazarenko_wave_2011, galtier_physics_2022}. Since the present system does not have a canonical Hamiltonian structure, the final results cannot be anticipated. Assuming a timescale separation between the linear and the nonlinear times, we introduce the interaction representation variable as
\begin{equation}
	c_{\kk} = \frac{\ak}{\epsilon} ~ e^{i \ok t},
\end{equation}
where the parameter $\epsilon \ll 1$ quantifies the strength of the nonlinearity. For stratified flows, $\epsilon$ corresponds to the Froude number $Fr$. Using (\ref{eq:WaveAction}), we obtain 
\begin{equation}
	\label{eq:InteractionRepresentation}
	\dot{c}_{\kk} = - i \epsilon \sum\limits_{1,2} ~ \Votk ~ \delta_{12}^{\kk} ~ c_1 c_2 ~ e^{i \omega_{12}^{\kk}t}  - 2 i \epsilon \sum\limits_{1,2} ~ \Votk ~ \delta_{\kk2}^1 ~ c_1 c_2^* ~ e^{- i \omega_{\kk2}^{1}t} - i \epsilon \sum\limits_{1,2} ~ \Votk ~ \delta_{\kk12} ~ c_1^* c_2^* ~ e^{- i \omega_{\kk 1 2}t},
\end{equation}
where $\omega_{12}^{\kk} \equiv \ok - \oo - \ot$ and $\omega_{\kk 1 2} \equiv - \ok - \oo - \ot$. The derivation of the kinetic equation describing the evolution of the wave action spectrum
\begin{equation}
	\label{eq:WaveActionSpectrum}
	n_{\kk} \equiv \left( \frac{2 \pi}{L} \right)^3 \left\langle |c_{\kk}|^2 \right\rangle
\end{equation}
in the infinite size limit ($L \rightarrow \infty$) and in the small nonlinearity limit ($\epsilon \rightarrow 0$) is given in Appendix~\ref{appendix1}. These computations follow the steps described in the book \citep{nazarenko_wave_2011}. The final result is 
\begin{align}
	\label{eq:KineticEquation}
	\dot{n}_{\kk} &= St_{\kk} = \int ~ \left[ \Rotk - \Qkto - \Qkot  \right] ~ \diff^3\kk_1 ~ \diff^3\kk_2, \\
	\label{eq:KineticEquationR}
	\Rotk &= 4 \pi \epsilon^2 ~ \delta(\kk - \kk_1 - \kk_2) ~ \delta(\omega_{12}^{\kk}) ~ \Votk ~ \left( \Votk ~ n_1 n_2 - \Vkot ~ n_{\kk} n_1 - \Vkto ~ n_{\kk} n_2 \right) + \textit{O}(\epsilon^3), \\
	\label{eq:KineticEquationQ}
	\Qkto &= 4 \pi \epsilon^2 ~ \delta(\kk_1 - \kk - \kk_2) ~ \delta(\omega_{\kk2}^{1}) ~ \Votk ~ \left(  \Vkto  ~ n_{\kk} n_2 - \Vkot ~ n_1 n_{\kk} - \Votk ~ n_1 n_2 \right) + \textit{O}(\epsilon^3).
\end{align}
It can be shown that this kinetic equation is equivalent to the one derived by \citet{caillol_kinetic_2000}. Equations (\ref{eq:KineticEquation}-\ref{eq:KineticEquationQ}) are relatively compact and more suitable for theoretical treatments. The total wave energy balance equation is
\begin{align}
	\dot{E} &= \int ~ \ok ~ \dot{n}_{\kk} ~ \diff^3\kk = \int ~ \ok ~ \left[ \Rotk - \Qkto - \Qkot  \right] ~ \diff^3\kk ~ \diff^3\kk_1 ~ \diff^3\kk_2 \\
	&= \int ~ \left[ \ok \Rotk - \oo \Qotk - \ot \Q_{21}^{\kk}  \right] ~ \diff^3\kk ~ \diff^3\kk_1 ~ \diff^3\kk_2 \\
	&=  4 \pi \epsilon^2 ~ \int ~ \delta(\kk - \kk_1 - \kk_2) ~ \delta(\omega_{12}^{\kk}) \\
	\nonumber
	&~~~~~~~~~~ \times  \left( \ok \Votk - \oo \Vkto - \ot \Vkot \right) ~ \left( \Votk ~ n_1 n_2 - \Vkot ~ n_{\kk} n_1 - \Vkto ~ n_{\kk} n_2 \right) ~ \diff^3\kk ~ \diff^3\kk_1 ~ \diff^3\kk_2,
\end{align}
which is zero because of the symmetry of the interaction coefficients ($\ref{eq:SymmetriesInteractionCoefficients}$). Note that it is not necessary to use the resonance condition in frequencies to prove that the kinetic equation conserves energy. $E$ is also an invariant of the Boussinesq equations (\ref{eq:Continuity}-\ref{eq:Buoyancy}) and of the wave action equation (\ref{eq:WaveAction}). \\

For now, the kinetic equation (\ref{eq:KineticEquation}-\ref{eq:KineticEquationQ}) does not take a standard form typical for canonical Hamiltonian systems, except if the relation $\Rotk = \Qotk$ holds. For this to happen, $\Votk$ should be fully symmetric with respect to the $3$ indices when evaluated on the resonant manifold $\kk = \kk_1 + \kk_2$, $\ok = \oo + \ot$ (or permutations) such that we could write $\Votk = \Vkto = \Vkot$ in $\Rotk$, $\Qkto$, and $\Qkot$. Such a symmetry could be expected since the kinetic equation obtained in the Lagrangian formalism is almost canonical \citep{hasselmann_nonlinear_1967, muller_dynamics_1975, olbers_nonlinear_1976}. As already shown in Figure~\ref{figure2}~(\textit{b}), resonance condition in wave vectors is not sufficient to have $\Votk = \Vkto = \Vkot$. However, adding the constraint of resonance condition in frequencies eventually allows to satisfy this symmetry, as will be explained in the subsection \ref{subsection3c}. 

\subsection{Steady, scale invariant spectra}
\label{subsection3b}

Despite its compact form, analytical solution to the kinetic equation (\ref{eq:KineticEquation}) are not easy to find. The only exception is for $n_{\kk} \propto 1/\ok$, corresponding to the equilibrium (RJ) spectrum with equipartition of energy. It is worth mentioning that this RJ spectrum can be obtained without using the resonance condition in frequencies, but only the symmetry of the interaction coefficients (\ref{eq:SymmetriesInteractionCoefficients}) when the wave vector resonance condition is satisfied. We can try to find other steady state solutions to the kinetic equation, in the non-hydrostatic case, by using the ansatz
\begin{equation}
	\label{eq:ScaleInvariantSpectrum}
	n_{\kk} = k^{\nu} f(\thk,\phk)
\end{equation}
corresponding to a separable, scale invariant spectrum. We adapt the computations of \citet{shavit_kinetic_2023} in order to find a possible value for the exponent $\nu$. We first write the evolution equation for the energy density averaged over angles $e(k,t)$. Namely,
\begin{equation}
	\label{eq:ShellEnergy}
	\dot{e}(k,t) = \int ~ \ok ~ \dot{n}_{\kk} ~ \sin \thk \diff \thk ~ \diff \phk = \int ~ \left[ \ok \Rotk - \ok \Qkto - \ok \Qkot  \right] ~ \diff^3\kk_1 ~ \diff^3\kk_2 ~ \sin \thk \diff \thk ~ \diff \phk.
\end{equation}
In steady state, the integral in the r.h.s of (\ref{eq:ShellEnergy}) must be zero. If we assume (\ref{eq:ScaleInvariantSpectrum}), this integral is also only a function of $k$, $\nu$, and depends on the angular variable via some function $f$. To find possible values for $\nu$, we adapt the Zakharov transformation, so we replace
\begin{equation}
	\label{eq:ZTIsotropic}
	k_1 \rightarrow \frac{k}{k_1} k, ~~~~ k_2 \rightarrow \frac{k}{k_1} k_2, ~~~~ (\thk,\phk) \leftrightarrow (\tho,\pho)
\end{equation}
in the integral with $\Qkto$, and a similar transformation for the integral with $\Qkot$ in (\ref{eq:ShellEnergy}). We then obtain the stationarity condition
\begin{align}
	\nonumber
	0 &= 4 \pi \epsilon^2 ~ \int ~ \delta(\kk - \kk_1 - \kk_2) ~ \delta(\omega_{12}^{\kk}) \\
	\label{eq:SteadyConditionIsotropic}
	& ~~~~~~~~~~ \times  \left[ \ok \Votk - \left( \frac{k_1}{k} \right)^{\chi} \oo \Vkto - \left( \frac{k_2}{k} \right)^{\chi} \ot \Vkot \right] \\
	\nonumber
	& ~~~~~~~~~~ \times \left( \Votk ~ n_1 n_2 - \Vkot ~ n_{\kk} n_1 - \Vkto ~ n_{\kk} n_2 \right) ~ \diff^3\kk_1 ~ \diff^3\kk_2 ~ \sin \thk \diff \thk ~ \diff \phk
\end{align}
where $\chi \equiv \alpha - 2 \beta - 2 \nu - 2 d$, with $d=3$ being the number of spatial dimensions. Condition (\ref{eq:SteadyConditionIsotropic}) is satisfied when $\chi = 0$ due to symmetry of the interaction coefficients (\ref{eq:SymmetriesInteractionCoefficients}). It leads to the exponent $\nu = -4$, which is consistent with the high frequency limit of the Garrett-Munk spectrum ($\propto k_h^{-4} |k_z|^{0}$), the PR spectrum ($\propto k_h^{-7/2} |k_z|^{-1/2}$), the spectrum obtained in the hydrostatic limit by considering induced-diffusion triads only \citep{lanchon_energy_2023} ($\propto k_h^{-3} |k_z|^{-1}$), and oceanic measurements \citep{lvov_energy_2004}. \\

Testing the validity of $\nu = -4$ is beyond the scope of this study. For this, we need to show that the collision integral converges in the vicinity of $\nu = -4$, for some yet unknown $f(\thk,\phk)$. This analysis has been done in the hydrostatic limit for bi-homogenous spectra $\propto k_h^{\nu_h} |k_z|^{\nu_z}$, and it has been found that the line $\nu = \nu_h + \nu_z = -4$ corresponds to non-physical spectra because of collision integral divergences \citep{lvov_oceanic_2010,dematteis_downscale_2021}. Yet, the divergence in the hydrostatic limit does not imply that $\nu=-4$ is unrealizable outside this limit because the convergence conditions could be less restrictive in the non-hydrostatic case. Moreover, the ansatz (\ref{eq:ScaleInvariantSpectrum}) is more general than $n_{\kk} = C k_h^{\nu_h} |k_z|^{\nu_z}$, which may allow other local spectra. Finally, it is important to note that the angular dependence $f(\thk,\phk)$ is embedded everywhere in the integrand in a nontrivial way and could lead to a cancellation of the collisional integral for $\nu \neq -4$. Indeed, the solution obtained by \citet{lvov_oceanic_2010,dematteis_downscale_2021} in the hydrostatic limit is of such a type.

\subsection{Canonical form and resonant manifold}
\label{subsection3c}

We proceed now to show that $\Votk = \Vkto = \Vkot$ on the resonant manifold, and thus that the kinetic equation \eqref{eq:WaveActionSpectrum} can be written in a canonical form. We first use the symmetry (\ref{eq:SymmetriesInteractionCoefficients}), together with the resonance condition in frequencies $\ok = \oo + \ot$, to readily show that 
\begin{equation}
	\label{eq:SymmetiesInteractionCoefficientsResonant}
	\delta(\kk-\kk_1-\kk_2) \delta(\omega_{12}^{\kk}) \left[ \oo (\Votk - \Vkto) + \ot (\Votk - \Vkot) \right] = 0.	
\end{equation}
Equation (\ref{eq:SymmetiesInteractionCoefficientsResonant}) has a simple geometrical meaning: the vectors $(\Votk - \Vkto, \Votk - \Vkot)$ and $(\oo, \ot=\ok-\oo)$ are orthogonal for all points of the resonant manifold. Because the $\ok$'s only depends on the angles $\thk$'s, to satisfy \eqref{eq:SymmetiesInteractionCoefficientsResonant} while varying  $k$ and $k_1$ (at fixed angles), the vectors $(\Votk - \Vkto, \Votk - \Vkot)$ must remain co-linear. The orthogonality condition (\ref{eq:SymmetiesInteractionCoefficientsResonant}) thus leads to
\begin{equation}
	\label{eq:OrthogonalityInteractionCoefficients}
	\left. \begin{pmatrix}
		\Votk - \Vkto \\ \Votk - \Vkot
	\end{pmatrix} \right|_{(k,k_1,\thk,\tho)} = g(k,k_1) 
	\left. \begin{pmatrix}
		\Votk - \Vkto \\ \Votk - \Vkot
	\end{pmatrix}\right|_{(1,1,\thk,\tho)} \perp 
	\begin{pmatrix}
		\oo \\ \ot
	\end{pmatrix} ~~~~ \forall (\oo, \ot),
\end{equation}
where $g$ is an unknown function that depends on $k$ and $k_1$ only. We have used here the fact that the resonant manifold can be parametrised using the variables $(k,k_1,\thk,\tho)$. Using equation (\ref{eq:OrthogonalityInteractionCoefficients}) in (\ref{eq:SymmetiesInteractionCoefficientsResonant}), we see that whether $g(k,k_1)=0$, or $\oo \left.(\Votk - \Vkto)\right|_{(1,1,\thk,\tho)} + \ot \left.(\Votk - \Vkot)\right|_{(1,1,\thk,\tho)}= 0$. Equation (\ref{eq:OrthogonalityInteractionCoefficients}) therefore allows us to reduce the analysis to $(k,k_1)=(1,1)$. Using the symbolic computational capabilities of the Mathematica software \citep{mathematica}, we show that 
\begin{align}
	\nonumber
	\left.\left( \Votk - \Vkto \right)\right|_{(1,1,\thk,\tho)} &= -\sqrt{\frac{N}{128}} \frac{1}{\cos(2\thk) + \cos(2\tho) + 4\sin\thk \sin\tho} \\
	\nonumber
	&~~ \times \left\{ \left[  \csc\left(\frac{\thk-\tho}{2}\right) \sin\left( \frac{\thk + \tho}{2}\right) \right. \right. \\ 
    \label{eq:Mathematica}
	&~~ \times (9 \cos(2\thk) - 3\cos(\thk-3\tho) - 
	8\cos(2\thk-2\tho)  \\
	\nonumber
	&~~~~~~ + 16\cos(\thk-\tho) \\
	\nonumber
	&~~~~~~- 3 (4 + \cos\left(3\thk-\tho\right) - 3\cos(2\tho) + 
	4\cos(\thk+\tho))) \\
	\nonumber
	&\left. \left. \left( \sqrt{-\csc\thk + \csc\tho} - 
	\csc\tho \sqrt{\sin\tho - \csc\thk \sin^2\tho} \right) \right] \right\} = 0.
\end{align}
It then follows from equations (\ref{eq:SymmetiesInteractionCoefficientsResonant}-\ref{eq:OrthogonalityInteractionCoefficients}) that 
\begin{equation}
	\label{eq:FullSymmetryInteractionCoefficients}
	\delta(\kk-\kk_1-\kk_2) \delta(\omega_{12}^{\kk}) \Votk = \delta(\kk-\kk_1-\kk_2) \delta(\omega_{12}^{\kk}) \Vkto = \delta(\kk-\kk_1-\kk_2) \delta(\omega_{12}^{\kk}) \Vkot,
\end{equation}
which is the desired symmetry to put the kinetic equation in a canonical form. \\

The kinetic equation (\ref{eq:KineticEquation}) can therefore be rewritten
\begin{align}
	\label{eq:KineticEquationCanonical}
	\dot{n}_{\kk} &= St_{\kk} = \int ~ \left[ \Rotk - \Rkto - \Rkot  \right] ~ \diff^3\kk_1 ~ \diff^3\kk_2, \\
	\label{eq:KineticEquationRCanonical}
	\Rotk &= 4 \pi \epsilon^2 ~ \delta(\kk - \kk_1 - \kk_2) ~ \delta(\omega_{12}^{\kk}) ~ |\Votk|^2 ~ \left( n_1 n_2 - n_{\kk} n_1 - n_{\kk} n_2 \right).
\end{align}
Similarly to Rossby waves \cite{nazarenko_wave_2011}, the wave action dynamics (\ref{eq:WaveAction}) does not have a canonical Hamiltonian structure, but the kinetic equation (\ref{eq:KineticEquationCanonical}) is the same as if the system were canonical because of the resonance condition in frequencies. Namely, we could have obtained equations (\ref{eq:KineticEquationCanonical}-\ref{eq:KineticEquationRCanonical}) by using the Hamilton equation $i\dot{a}_{\kk} = \frac{\delta H_{\rm{eff}}}{\delta a_{\kk}^*}$ together with the effective Hamiltonian $H_{\rm{eff}} = \frac{1}{2} \sum\limits_3 ~ \omega_3 |a_3|^3 + \sum\limits_{1,2,3} ~ \left( \delta_{12}^3 V_{12}^3 a_1 a_2 a_3^* + \delta_{23}^1 V_{23}^1 a_1^* a_2 a_3^* \right)$. Yet, it is not equivalent to the original wave mode equation (\ref{eq:WaveAction}) outside the resonant manifold. \\

In the case of a wave action spectrum invariant under rotation around the stratification axis, i.e. $n_{\kk}=n(k_h,k_z,t)$, the kinetic equation (\ref{eq:KineticEquationCanonical}-\ref{eq:KineticEquationRCanonical}) takes a simpler form after integrating over azimuthal angles $(\pho,\pht)$:
\begin{align}
	\label{eq:KineticEquationHorizontal}
	\dot{n}_{\kk} &= St_{\kk}^{(h)} = \int ~ \left[ \Rotkh - \Rktoh - \Rkoth  \right] ~ k_{1h} k_{2h} ~ \diff k_{1h} \diff k_{1z} ~ \diff k_{2h} \diff k_{2z}, \\
	\label{eq:KineticEquationRHorizontal}
	\Rotkh &= 8 \pi \epsilon^2 ~ \delta(k_z - k_{1z} - k_{2z}) ~ \delta(\omega_{12}^{\kk}) ~ \frac{|\Votk|^2}{\Delta} ~ \left( n_1 n_2 - n_{\kk} n_1 - n_{\kk} n_2 \right), \\
	\label{eq:Deltah}
	\Delta &= \frac{1}{2} \sqrt{(- k_h + k_{1h} + k_{2h}) (k_h - k_{1h} + k_{2h}) (k_h + k_{1h} - k_{2h}) (k_h + k_{1h} + k_{2h})}.
\end{align}
In the latter equations, the interaction coefficients are evaluated using $\kk_h = \kk_{1h} + \kk_{2h}$, and similar relations obtained by permutations of wave vectors. It follows that they are functions of $(k_h, k_z, k_{1h}, k_{1z}, k_{2h}, k_{2z})$ only. The factor 2 arising in $\Rotkh$ (\ref{eq:KineticEquationRHorizontal}) when compared to $\Rotk$ (\ref{eq:KineticEquationRCanonical}) comes from the fact that there are two solutions $(\pho,\pht)$ to $\kk_h = \kk_{1h} + \kk_{2h}$ (or permutations of wave vectors) for each $(k_h,k_z,k_{1h},k_{1z},k_{2h},k_{2z})$. $\Delta$ is the area of the triangle formed by the horizontal wave vectors. It arises from the average of $\delta(\kk_h - \kk_{1h} - \kk_{2h})$.
We can further simplify the collision integral (\ref{eq:KineticEquationHorizontal}) by working with polar coordinates $(k_h,k_z)=k(\sin \thk, \cos \thk)$, $(k_{1h},k_{1z})=k_1(\sin \tho, \cos \tho)$, and $(k_{2h},k_{2z})=k_2(\sin \tht, \cos \tht)$. Because the last two terms of the collision integral are symmetric with respect to $\kk_1 \leftrightarrow \kk_2$, we obtain
\begin{align}
	St_{\kk}^{(h)} &= \I_{\kk} - 2 \J_{\kk}, \\
	\label{eq:Ik}
	\I_{\kk} &= 8 \pi \epsilon^2 ~ \int ~ \delta(k \cos \thk - k_1 \cos \tho - k_2 \cos \tht) ~ \delta(N(\sin \thk - \sin \tho - \sin \tht))  \\
	\nonumber 
	&~~~~~~~~ \times  \frac{|\Votk|^2}{\Delta} ~ \left( n_1 n_2 - n_{\kk} n_1 - n_{\kk} n_2 \right) ~ k_1^2 \sin \tho ~ k_2^2 \sin \tht ~ \diff \tho \diff k_1 \diff \tht \diff k_2, \\
	\label{eq:Jk}
	\J_{\kk} &= 8 \pi \epsilon^2 ~ \int ~ \delta(k_1 \cos \tho - k \cos \thk - k_2 \cos \tht) ~ \delta(N(\sin \tho - \sin \thk - \sin \tht))  \\
	\nonumber 
	&~~~~~~~~ \times  \frac{|\Votk|^2}{\Delta} ~ \left(  n_{\kk} n_2 - n_1 n_{\kk} - n_1 n_2 \right) ~ k_1^2 \sin \tho ~ k_2^2 \sin \tht ~ \diff \tho \diff k_1 \diff \tht \diff k_2.
\end{align}

\begin{figure}
	\centerline{\includegraphics[width=1\textwidth]{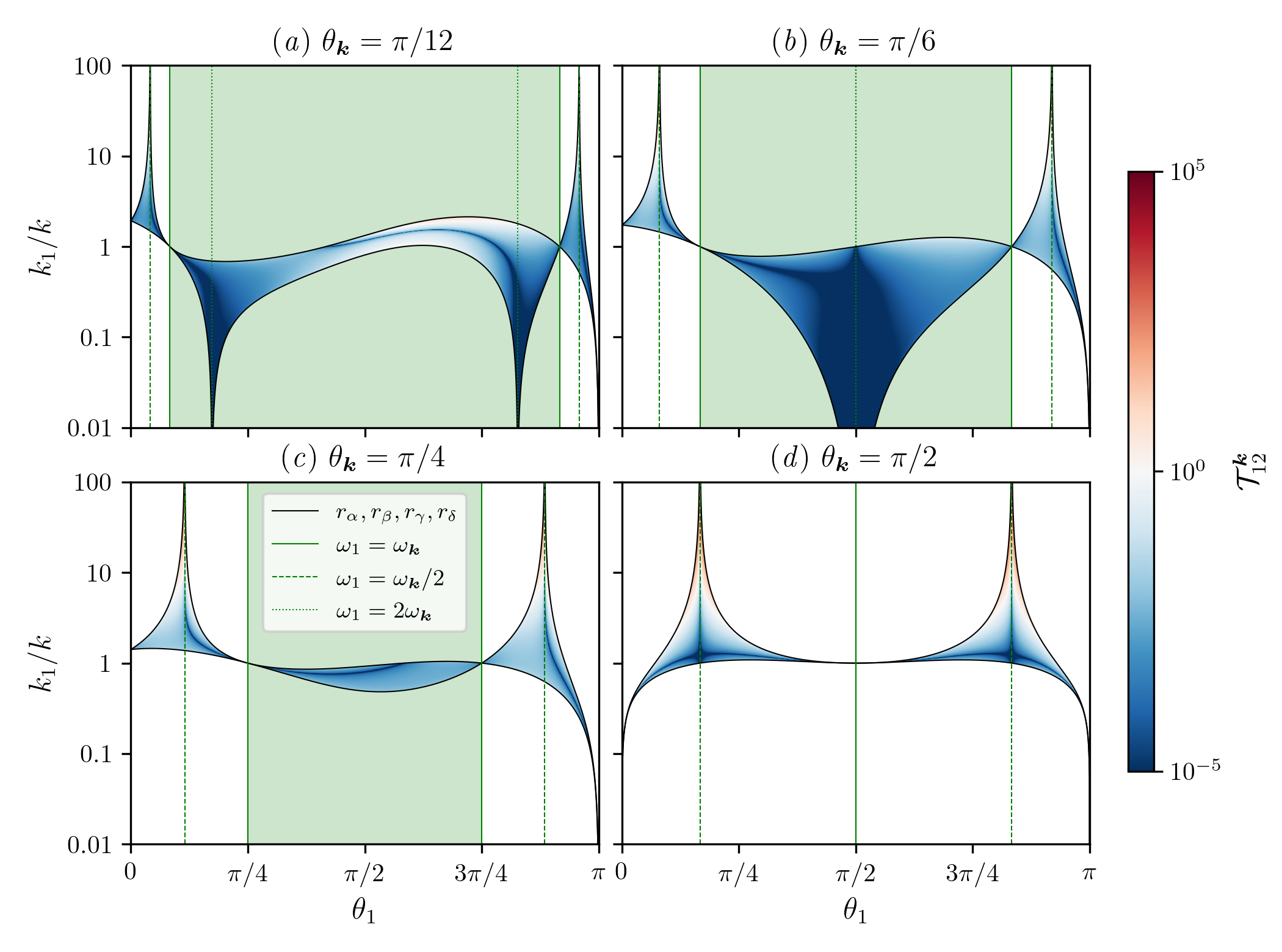}}
		\caption{Transfer coefficient on the integration domain of the collisional integral for several $\thk \in [0;\pi/2]$: (\textit{a})~$\thk=\pi/12$, (\textit{b})~$\thk=\pi/6$, (\textit{c})~$\thk=\pi/4$, and (\textit{d})~$\thk=\pi/2$. The borders of the domains are given by the critical lines $\ra$, $\rb$, $\rc$, and $\rd$ (\ref{eq:xa}-\ref{eq:xd}). $\D_{\I}$ lies outside $\oo = \ok$ lines (white background), while $\D_{\J}$ is contained inside $\oo = \ok$ lines (green background).}
		\label{figure3}	
\end{figure}

To go further, we need to parametrize the resonant manifold, corresponding to the set of wave vectors $(\kk, \kk_1, \kk_2)$ satisfying the resonant conditions
\begin{equation}
	\label{eq:ResonantManifoldEquations}
	\begin{cases}
		\kk = \kk_1 + \kk_2 \\
		\ok = \oo + \ot
	\end{cases}
	~~ \text{or} ~~
	\begin{cases}
		\kk_1 = \kk + \kk_2 \\
		\oo = \ok + \ot
	\end{cases}.
\end{equation}
It appears that it is relatively easy to find $(k_{2h}, k_{2z})$ as a function of $(k_h, k_z, k_{1h}, k_{1z})$. This leads to 
\begin{equation}
	\label{eq:ResonantManifold}
	\begin{cases}
		k_{2z} = k_z - k_{1z} \\
		k_{2h} = |k_z - k_{1z}||\tan \tht|
	\end{cases}
	~~ \text{or} ~~
	\begin{cases}
		k_{2z} = k_{1z} - k_z \\
		k_{2h} = |k_{1z} - k_z||\tan \tht|
	\end{cases}
\end{equation}
with $|\tan \tht| = \frac{|\sin \thk - \sin \tho|}{\sqrt{1-(\sin \thk - \sin \tho)^2}}$ and $\sin \tho = k_{1h}/k_1$. These solutions are valid if and only if
\begin{equation}
	\label{eq:DIDJ}	
	\begin{cases}
		\ok \geq \oo \\
		k_h \leq k_{1h} + k_{2h} \\
		k_{1h} \leq k_{2h} + k_h \\
		k_{2h} \leq k_h + k_{1h}
	\end{cases}
	~~ \text{or} ~~
	\begin{cases}
		\oo \geq \ok \\
		k_{1h} \leq k_h + k_{2h} \\
		k_h \leq k_{2h} + k_{1h} \\
		k_{2h} \leq k_{1h} + k_h
	\end{cases}
\end{equation}
otherwise there is no solution. We can use the $\delta$-Dirac function to perform integration over $k_{2h}$ and $k_{2z}$ such that we finally obtain 
\begin{align}
	\label{eq:IkSimple}
	\I_{\kk} &= \frac{8 \pi \epsilon^2}{N} ~ \int\limits_{\D_{\I}} ~ \frac{k_1^2 \sin \tho ~ k_2^2 \sin \tht}{\cos^2 \tht ~ \Delta} ~ |\Votk|^2 ~ \left( n_1 n_2 - n_{\kk} n_1 - n_{\kk} n_2 \right) ~ \diff \tho \diff k_1, \\
	\label{eq:JkSimple}
	\J_{\kk} &= \frac{8 \pi \epsilon^2}{N} ~ \int\limits_{\D_{\J}} ~ \frac{k_1^2 \sin \tho ~ k_2^2 \sin \tht}{\cos^2 \tht ~ \Delta} ~ |\Votk|^2 ~ \left( n_{\kk} n_2 - n_1 n_{\kk} - n_1 n_2 \right) ~ \diff \tho \diff k_1,
\end{align}
where $\D_{\I}$ and $\D_{\J}$ are the integration domains of the resonant manifold given by conditions (\ref{eq:DIDJ}) detailed later and shown on Figure~\ref{figure3}. Note that it is possible to parameterize the resonant manifold with the variables $(k_1,k_2)$ instead of $(k_1,\tho)$. However, it requires solving a polynomial equation of order 4 to find the angles $(\tho,\tht)$. Despite the fact that it is mathematically doable, it leads to equations that are much more difficult to use than the one obtained when we parameterize the resonant manifold using $(k_1,\tho)$. \\

For simplicity, in order to define analytically $\D_{\I}$ and $\D_{\J}$, we will assume that the wave action spectrum is invariant under the transformation $k_z \rightarrow - k_z$, i.e. $n(k_h, k_z) = n(k_h, -k_z)$. In this way, we can restrict the study of the collision integral to $k_z \geq 0$ (i.e., $0 \leq \thk \leq \pi/2$). The study of the resonant surface when the wave action spectrum does not have this symmetry requires considering the case with $k_z \leq 0$ (i.e., $\pi/2 \leq \thk \leq \pi$), which is longer but technically not more difficult. The borders of the integration domains $\D_{\I}$ and $\D_{\J}$ correspond to the zeros of $\Delta$ given by (\ref{eq:Deltah}). Note that the borders of the resonant manifold are attained for wave triads with co-linear horizontal projection, i.e. waves on the same vertical plane. They are easily obtained in the variable $k_1/k$ as a function of $(\thk,\tho)$. For each $(\thk,\tho)$, the zeros of $\Delta$ are attained on two of the following lines
\begin{align}
	\label{eq:xa}
	\frac{k_1}{k} = \ra(\thk, \tho) &\equiv \frac{\sqrt{1-(\sin \thk - \sin \tho)^2}\sin(\thk)-\sin(\tho)\cos(\thk)+\sin(2\thk)/2}{\sqrt{1-(\sin \thk - \sin \tho)^2}\sin(\tho)+\sin(\thk)\cos(\tho)-\sin(2\tho)/2} \\ 
	\frac{k_1}{k} = \rb(\thk, \tho) &\equiv \frac{\sqrt{1-(\sin \thk - \sin \tho)^2}\sin(\thk)+\sin(\tho)\cos(\thk)-\sin(2\thk)/2}{\sqrt{1-(\sin \thk - \sin \tho)^2}\sin(\tho)-\sin(\thk)\cos(\tho)+\sin(2\tho)/2} \\
	\frac{k_1}{k} = \rc(\thk, \tho) &\equiv \frac{-\sqrt{1-(\sin \thk - \sin \tho)^2}\sin(\thk)-\sin(\tho)\cos(\thk)+\sin(2\thk)/2}{\sqrt{1-(\sin \thk - \sin \tho)^2}\sin(\tho)+\sin(\thk)\cos(\tho)-\sin(2\tho)/2} \\
	\label{eq:xd}
	\frac{k_1}{k} = \rd(\thk, \tho) &\equiv \frac{-\sqrt{1-(\sin \thk - \sin \tho)^2}\sin(\thk)+\sin(\tho)\cos(\thk)-\sin(2\thk)/2}{\sqrt{1-(\sin \thk - \sin \tho)^2}\sin(\tho)-\sin(\thk)\cos(\tho)+\sin(2\tho)/2}
\end{align}
leading to the subdomains of $\D_{\I}$ and $\D_{\J}$ that are listed in Table \ref{tab:IntegrationDomains}. It is worth mentioning that the description of the resonant manifold of internal waves (accounting for rotation) in the $(k_1/k, \thk, \tho)$ variables is available in \citet{olbers_energy_1974} (see section 4 of this reference). \\

\begin{table}
	\begin{center}
	\def~{\hphantom{0}}
	\begin{tabular}{|c|c|c|c|c|c|}
		\hline
		\multicolumn{2}{|c|}{$\D_{\I}$} & \multicolumn{2}{c|}{$\D_{\J}$ ~ $(\thk \leq \pi/6)$} &
		\multicolumn{2}{c|}{$\D_{\J}$ ~ $(\thk \geq \pi/6)$} \\
		\hline
		$\tho$ & $k_1/k$ & $\tho$ & $k_1/k$ & $\tho$ & $k_1/k$ \\
		\hline
		$\left[0; \arcsin \left( \frac{\sin \thk}{2} \right)\right]$ & $\left[ \ra; \rd \right]$ & $\left[ \thk;\arcsin \left( 2 \sin \thk \right) \right]$ & $\left[ \ra;\rb \right]$ & $\left[\thk;\pi - \thk \right]$ & $\left[ \ra;\rb \right]$ \\
		$\left[ \arcsin \left( \frac{\sin(\thk)}{2} \right); \thk \right]$ & $\left[ \ra; \rb \right]$ & $\left[ \arcsin \left( 2 \sin \thk \right); \pi - \arcsin \left( 2 \sin \thk \right) \right]$  &
		$\left[ \rd;\rb \right]$ & & \\
		$\left[ \pi - \thk; \pi - \arcsin \left( \frac{\sin(\thk)}{2} \right)\right]$  &
		$\left[ \rb; \ra \right]$ & $\left[\pi - \arcsin \left( 2 \sin \thk \right);\pi - \thk \right]$ &
		$\left[ \ra;\rb \right]$ & & \\
		$\left[ \pi - \arcsin \left( \frac{\sin(\thk)}{2} \right); \pi \right]$ & $\left[ \rb; \rc \right]$  & & & & \\
		\hline
	\end{tabular}
	\caption{Subdomains of the resonant manifolds $\D_{\I}$ and $\D_{\J}$ for $\thk \in \left[0;\pi/2\right]$.}
	\label{tab:IntegrationDomains}
	\end{center}
\end{table}

In Figure~\ref{figure3}, we represent the transfer coefficient $\Totk \equiv k_1^2 \sin \tho k_2^2 \sin \tht |\Votk|^2 / (\cos^2 \tht \Delta)$ on the integrations domains $\D_{\I}$ and $\D_{\J}$. $\Totk$ measures the strength of the interaction of the triad (\ref{eq:IkSimple}-\ref{eq:JkSimple}), which is important information for studying the evolution of IGW. As can be expected from earlier studies \citep{mccomas_bretherton_resonant_1977, muller_nonlinear_1986, lvov_oceanic_2010, eden_numerical_2019, olbers_psi_2020, lanchon_energy_2023}, the transfer coefficient is important for triads with large scale separation, i.e., when $k_1$ or $k_2 \gg k$. The transfer coefficient is also large for triads with $k \sim k_1 \sim k_2$ at the border of the domain, corresponding to triads contained in a vertical plane. It is in line with \citet{dematteis_downscale_2021}, who showed that the dominant contribution to the collision integral for the steady state spectra, in the hydrostatic limit, is due to the horizontally co-linear wave triads. It has also been shown that local interactions correspond to an important part of the energy transfers for the Garrett-Munk spectrum \citep{wu_energy_2023}. \\

\subsection{Hydrostatic limit}
\label{subsection3d}

In strongly stratified flows, the energy tends to concentrate in modes with wave vectors such that the vertical component is much larger than the horizontal one. It is therefore worth considering the almost vertical propagation hypothesis (or hydrostatic limit) with $|k_z| \gg k_h$, and so $\ok = N k_h/k \simeq N k_h/|k_z|$. In that limit, our kinetic equation is equivalent to other kinetic equations of internal gravity wave turbulence \citep{muller_dynamics_1975, caillol_kinetic_2000, lvov_hamiltonian_2001}, as shown in the review of \citet{lvov_resonant_2012}. We have checked that our interaction coefficients are equal to the one of \citet{lvov_hamiltonian_2001} up to machine precision in the hydrostatic limit, when evaluated on the resonant manifold. In the hydrostatic limit, we can find stationary, axisymmetric, bi-homogeneous solutions $n_{\kk} = n(k_h,k_z) \propto k_h^{\nu_h} ~ |k_z|^{\nu_z}$ to the wave kinetic equation. This theoretical and numerical work has already been achieved in earlier studies \citep{pelinovsky_raevsky_weak_1977, caillol_kinetic_2000, lvov_hamiltonian_2001, lvov_energy_2004,lvov_oceanic_2010,dematteis_downscale_2021}, so we will not repeat all the computations here. Instead, we show that the KZ spectrum can be obtained without involving resonance condition in frequencies.  \\

In the hydrostatic limit, we have the following simplifications
\begin{align}
	&\eepo \cdot \kk_2 = \frac{k_{1z}}{k_1 k_{1h}} \kk_{1h} \cdot \kk_{2h} - \frac{k_{1h} k_{2z}}{k_1} \simeq \frac{k_{1z}}{|k_{1z}| k_{1h}} \kk_{1h} \cdot \kk_{2h} - \frac{k_{1h} k_{2z}}{|k_{1z}|}, \\
	&\eepo \cdot \eept = \frac{1}{k_1 k_2} \left( k_{1z} k_{2z} \frac{\kk_{1h} \cdot \kk_{2h}}{k_{1h} k_{2h}} + k_{1h} k_{2h} \right) \simeq \frac{k_{1z} k_{2z}}{|k_{1z}| |k_{2z}|} \frac{\kk_{1h} \cdot \kk_{2h}}{k_{1h} k_{2h}}
\end{align} 
and similar relations after permutations of wave vectors. Note that the scalar products of horizontal wave vectors (e.g., $\kk_{1h} \cdot \kk_{2h}$) are determined by the resonance condition for horizontal wave vectors, and are only functions of $(k_h,k_{1h},k_{2h})$. Outside the hydrostatic limit, the wave frequency and interaction coefficients are homogeneous in wave vector amplitudes $(k,k_1,k_2)$, but not in $(k_h,k_{1h},k_{2h})$ and $(|k_z|,|k_{1z}|,|k_{2z}|)$ separately. Contrarily, in the hydrostatic limit, the wave frequency and interaction coefficients are bi-homogeneous, i.e. are homogeneous in $(k_h,k_{1h},k_{2h})$ and $(|k_z|,|k_{1z}|,|k_{2z}|)$ separately. More precisely, the transformation $(k_h,k_{1h},k_{2h}) \rightarrow \mu_h (k_h,k_{1h},k_{2h})$, $(k_z,k_{1z},k_{2z}) \rightarrow \mu_z (k_z,k_{1z},k_{2z})$ changes the frequency and interaction coefficients in the following way
\begin{equation}
	\ok \rightarrow \bmu^{\balpha} ~ \ok ~~~~ \text{and} ~~~~ \Votk \rightarrow \bmu^{\bbeta} ~ \Votk
\end{equation}
where we have used the notation of \citep{balk_physical_1990} $\bmu^{\balpha} \equiv \mu_h^{\alpha_h} \mu_z^{\alpha_z}$, with $\balpha = (1,-1)$ and $\bbeta = \left( \frac{3}{2}, -\frac{1}{2} \right)$. The homogeneity degrees inside and outside the hydrostatic limit are linked by $\alpha_h+\alpha_z=\alpha=0$ and $\beta_h+\beta_z=\beta=1$. \\

 We now perform the Zakharov-Kuznetsov transformation \citep{kuznetsov_turbulence_1972} for bi-homogenous spectra
\begin{equation}
	\label{eq:ZT}
	k_{1h} \rightarrow \frac{k_h}{k_{1h}} k_h, ~~~~ k_{2h} \rightarrow \frac{k_h}{k_{1h}} k_{2h}, ~~~~ k_{1z} \rightarrow \frac{k_z}{k_{1z}} k_z, ~~~~ k_{2z} \rightarrow \frac{k_z}{k_{1z}} k_{2z}
\end{equation}
to the integral with $\Qkto$, and a similar transformation for the integral with $\Qkot$ in the wave kinetic equation (\ref{eq:KineticEquation}). After the Zakharov-Kuznetsov transformation (\ref{eq:ZT}), the collision integral becomes 

\begin{align}
	\label{eq:CollisionIntegralKZ}
	St_{\kk} &\rightarrow 4 \pi \epsilon^2 ~ \int ~ \delta(\kk - \kk_1 - \kk_2) ~ \delta(\omega_{12}^{\kk}) \\
	\nonumber
	&\times ~ \left[ \Votk - \Vkto ~ \left( \frac{\kk_1}{\kk} \right)^{\bchi} - \Vkot ~ \left( \frac{\kk_2}{\kk} \right)^{\bchi} \right] ~ \left( \Votk ~ n_1 n_2 - \Vkto ~ n_{\kk} n_2 - \Vkot ~ n_{\kk} n_1 \right) ~ \diff \kk_1 \diff \kk_2
\end{align}
with $\bd = (2,1)$ and $\bchi = \balpha - 2 \bbeta - 2 \bnu - 2 \bd$. The integrand in (\ref{eq:CollisionIntegralKZ}) is zero when $-\bnu = \balpha$ ($n_{\kk} \propto 1/\ok$), which corresponds to the RJ spectrum, and when $\bchi = \balpha$, which leads to the KZ spectrum $n_{\kk} \propto k_h^{-7/2} |k_z|^{-1/2}$. Interestingly, we obtain the RJ and the KZ spectra by using the symmetry of the interaction coefficients (\ref{eq:SymmetriesInteractionCoefficients}), and not the resonance on frequencies, which is unusual in WWT theory. As said in the introduction, the same theoretical spectrum was first obtained by \citet{pelinovsky_raevsky_weak_1977}, and later by \citet{caillol_kinetic_2000} and \citet{lvov_hamiltonian_2001} using different formalisms. Yet, the Zakharov-Kuznetsov transformation (\ref{eq:ZT}) transformation is a non-identity transformation that takes the limit of zero wave numbers to infinity and vice versa, which may lead to the cancellation of oppositely signed divergences. If the original collision integral converges and is equal to zero, then the found spectrum is indeed a valid solution, in which case the spectrum is called local. If not, then the spectrum in question is a spurious solution; it is called a non-local spectrum \citep{nazarenko_wave_2011} and is not physically realizable. It turned out that the KZ (or PR) spectrum is non-local \citep{caillol_kinetic_2000}.  \\

The collision integral is expected to have other zeros. This situation is typical for anisotropic media \citep{kuznetsov_turbulence_1972, balk_physical_1990, lvov_energy_2004}. To find these spectra, we need to compute the collision integral to find its zeros in the $(\nu_h,\nu_z)$ plane. It is important to note that the integrand in the collision integral can diverge if $k_{1h}$ or $k_{2h} \rightarrow 0$ which corresponds to Infrared (IR) divergence, and when $k_{1h}, k_{2h} \rightarrow \infty$ which correspond to Ultraviolet (UV) divergence. Depending on $(\nu_h,\nu_z)$, the divergences can be integrable (in which case the spectrum is local), or non-integrable (in which case the spectrum is non-local). To identify these divergences, we need to analyse the terms of the collision integral and see for which values of $(\nu_h,\nu_z)$ they lead to non-integrable divergences. The locality conditions have been obtained by \citet{lvov_oceanic_2010}, and it was shown that another steady state solution exists due to the opposite signs of IR and UV divergences. More precisely, the collision integral converges only on the segment $\nu_z = 0$, $-4 < \nu_h < -3$ and the steady-state spectrum $n_{\kk} \propto k_h^{-3.69}$ has been obtained by finding the zero of the collision integral on this segment numerically \citep{lvov_oceanic_2010,dematteis_downscale_2021}. Yet, any spectrum with $\nu_z \neq 0$ around that solution would lead to divergent collisional integral. It means that the collision integral is not differentiable with respect to $\nu_z$ and $\nu_h$ for spectra with $\nu_z=0$ and, therefore, the energy flux integrals are divergent (see \citet{zakharov_kolmogorov_1992} section 3.32). This rules out realisability of these spectra. Since our formalism is equivalent to \citep{lvov_hamiltonian_2001} in the hydrostatic limit, the same results are expected. \\

\section{Discussions and conclusions}
\label{section4}

We have presented a new derivation of the kinetic equation describing the weak internal gravity wave turbulence using poloidal-toroidal-shear decomposition \citep{craya_contribution_1957,herring_approach_1974,godeferd_toroidal_2010}. This decomposition is particularly well adapted to the problem because it offers a complete basis of the flow modes, uses standard Eulerian coordinates, and the poloidal velocity is the kinetic part of the wave mode coupled to the buoyancy. The resulting kinetic equation has an advantage to hold in the non-hydrostatic case. It is equivalent to the one obtained by \citet{caillol_kinetic_2000}, but is considerably more compact than the latter. The interaction coefficients satisfy symmetries \citep{remmel_new_2009} that impose energy conservation both at the wave action equation level, and at the kinetic equation level. It results that energy conservation in the kinetic equation can be demonstrated without using the resonance condition in frequencies, which is unusual in Weak Wave Turbulence theory. Similarly, we show that Rayleigh-Jeans and Kolmogorov-Zakharov spectra can be obtained using this symmetry, without involving resonance condition in frequencies. Adapting the computations of \citet{shavit_kinetic_2023}, we obtain a stationarity condition for a scale invariant spectra  $n_{\kk} = k^{\nu} f(\thk,\phk)$, and showed that $\nu=-4$ is a good candidate, in reasonably good agreement with known theoretical results and oceanic measurements \citep{lvov_energy_2004}. It is worth mentioning that in the hydrostatic limit, these spectra are known to lead to divergences of the collisional integral \citep{lvov_oceanic_2010,dematteis_downscale_2021}. Validity of a scale invariant spectra of the form $n_{\kk} = k^{-4} f(\thk,\phk)$ remains to be checked. We have shown that the interaction coefficients are fully symmetric with respect to permutation of wave-vectors on the resonant manifold, even outside the hydrostatic limit. It follows that the kinetic equation has a canonical structure, despite the fact that the original equation does not have a canonical Hamiltonian structure. Namely, the kinetic equation reads
\begin{align*}
	\dot{n}_{\kk} &= St_{\kk} = \int ~ \left[ \Rotk - \Rkto - \Rkot  \right] ~ \diff^3\kk_1 ~ \diff^3\kk_2, \\
	\Rotk &= 4 \pi \epsilon^2 ~ \delta(\kk - \kk_1 - \kk_2) ~ \delta(\omega_{12}^{\kk}) ~ |\Votk|^2 ~ \left( n_1 n_2 - n_{\kk} n_1 - n_{\kk} n_2 \right),
\end{align*}
with $\ok = N \sin \thk$, and interaction coefficients $\Votk$ given by equation (\ref{eq:InteractionCoefficients}). We have parametrised the resonant manifold, allowing us to give a simplified version of the kinetic equation for axisymmetric spectra. We have computed numerically the transfer coefficient, quantifying the strength of the interaction, for all triads of the resonant manifold. Consistently with earlier studies, we find that interactions corresponding to large scale separation \citep{mccomas_bretherton_resonant_1977,muller_nonlinear_1986,lvov_oceanic_2010,eden_numerical_2019,olbers_psi_2020,lanchon_energy_2023} and local interactions \citep{dematteis_downscale_2021, wu_energy_2023} both play an important role on the dynamics of IGW turbulence. In the hydrostatic limit, our kinetic equation is equivalent to many other formalisms \citep{lvov_hamiltonian_2001, lvov_resonant_2012} so we refer the reader to the numerous studies available in the literature, in particular \citet{lvov_oceanic_2010,dematteis_downscale_2021}.

\backsection[Supplementary data]{\label{SupMat}We provide the Mathematica script allowing us to prove analytically equation \eqref{eq:Mathematica}, and a Jupyter notebook that allows us to check the symmetry of the interaction coefficients on the resonant manifold by direct numerical computations.} 

\backsection[Acknowledgements]{We thank Jalal Shatah, Oliver Bühler, Michal Shavit, and Miguel Onorato for fruitful discussions. We are grateful to three referees whose remarks helped to significantly improve this work.}

\backsection[Funding]{This work was supported by the Simons Foundation through Grants No. 651471 GK and No. 651461 PPC.}

\backsection[Declaration of interests]{The authors report no conflict of interest.}

\backsection[Data availability statement]{No data is associated with this work.}

\backsection[Author ORCIDs]{\\
V. Labarre \href{https://orcid.org/0000-0002-5081-4008}{https://orcid.org/0000-0002-5081-4008}; \\
N. Lanchon \href{https://orcid.org/0000-0002-3931-7334}{https://orcid.org/0000-0002-3931-7334}; \\
P.-P. Cortet \href{https://orcid.org/0000-0002-0444-0906}{https://orcid.org/0000-0002-0444-0906}; \\
G. Krstulovic \href{https://orcid.org/0000-0002-9934-6292}{https://orcid.org/0000-0002-9934-6292}; \\
S. Nazarenko \href{https://orcid.org/0000-0002-8614-4907}{https://orcid.org/0000-0002-8614-4907} \\
}

\backsection[Author contributions]{
VL, GK, and SN have performed the analytical derivations. All the authors contributed to prove the canonical form of the kinetic equation and contributed to write the paper.
}

\appendix

\section{Derivation of the wave kinetic equation}
\label{appendix1}

Here, we derive the wave kinetic equation (\ref{eq:KineticEquation}-\ref{eq:KineticEquationQ}) starting from the interaction representation variable equation (\ref{eq:InteractionRepresentation}). Firstly, we remark that the last term will lead to interaction between modes satisfying $\ok + \oo + \ot = 0$ after taking the limit $\epsilon \rightarrow 0$. Since $\ok \geq 0$, this term will therefore correspond to interactions between shear modes, which are not taken into account here. For this reason, we do not need to consider the last term of the equation (\ref{eq:InteractionRepresentation}) to obtain the wave kinetic equation. The next step is to consider an intermediate (between linear and nonlinear) time $T$, $\frac{2\pi}{\ok} \ll T = \frac{2\pi}{\epsilon \ok} \ll \frac{2\pi}{\epsilon^2 \ok}$, use an expansion in $\epsilon$ up to second order, $c_{\kk}(T) = c_{\kk}^{(0)} + \epsilon c_{\kk}^{(1)} + \epsilon^2 c_{\kk}^{(2)} + \textit{O}(\epsilon^3)$, and obtain the following expression for the expansion of $c_{\kk}(T)$:
\begin{align}
	\label{eq:InteractionVariableExpansionStart}
	c_{\kk}^{(0)}(T) &= c_{\kk}^{(0)}(0) = c_{\kk}(0), \\
	c_{\kk}^{(1)}(T) &= - i \sum\limits_{1,2} ~ \Votk ~ \delta_{12}^{\kk} ~ c_1^{(0)}  c_2^{(0)} ~ \Gamma_T(\omega_{12}^{\kk}) - 2 i \sum\limits_{1,2} ~ \Votk ~ \delta_{\kk2}^1 ~ c_1^{(0)} c_2^{(0)*} ~ \Gamma_T^*(\omega_{\kk2}^{1}), \\
	\label{eq:InteractionVariableExpansionEnd}
	\nonumber
	c_{\kk}^{(2)}(T) &= - 2 \sum\limits_{1,2,3,4} ~ \Votk ~ \delta_{12}^{\kk} ~ \left[ V_{34}^2 ~ \delta_{34}^2 ~ c_1^{(0)} c_3^{(0)} c_4^{(0)} ~ \Lambda_T(\omega_{34}^2, \omega_{12}^{\kk}) \right. \\
	\nonumber
	&\left. \hspace{3cm} + 2 V_{34}^2 ~ \delta_{24}^3 ~ c_1^{(0)} c_3^{(0)} c_4^{(0)*} ~ \Lambda_T(-\omega_{24}^3, \omega_{12}^{\kk})  \right] \\
	& ~~ - 2 \sum\limits_{1,2,3,4} ~ \Votk ~ \delta_{\kk2}^1 ~ \left[ V_{34}^1 ~ \delta_{34}^1 ~ c_2^{(0)*} c_3^{(0)} c_4^{(0)} ~ \Lambda_T(\omega_{34}^1, -\omega_{\kk2}^{1}) \right. \\ 
	\nonumber
	&\left. \hspace{3cm} + 2 V_{34}^1 ~ \delta_{14}^3 ~ c_2^{(0)*} c_3^{(0)} c_4^{(0)*} ~ \Lambda_T(-\omega_{14}^3, -\omega_{\kk2}^{1})  \right] \\
	\nonumber
	& ~~ + 2 \sum\limits_{1,2,3,4} ~ \Votk ~ \delta_{\kk2}^1 ~ \left[ V_{34}^2 ~ \delta_{34}^2 ~ c_1^{(0)} c_3^{(0)*} c_4^{(0)*} ~ \Lambda_T(-\omega_{34}^2, -\omega_{\kk2}^{1}) \right. \\
	\nonumber
	&\left. \hspace{3cm} + 2 V_{34}^2 ~ \delta_{24}^3 ~ c_1^{(0)} c_3^{(0)*} c_4^{(0)} ~ \Lambda_T(\omega_{24}^3, -\omega_{\kk2}^{1})  \right] 
\end{align}
where $\Gamma_T(x) \equiv \int\limits_{0}^{T} ~ e^{ixt} ~ \diff t$ and $\Lambda_T(x,y) \equiv \int\limits_{0}^{T} ~ \Gamma_T(x) ~ e^{iyt} ~ \diff t$. We need to compute 
\begin{equation}
	\label{eq:WaveActionExpansion}
	\left\langle \left| c_{\kk} \right|^2 \right\rangle = \left\langle \left| c_{\kk}^{(0)} \right|^2 \right\rangle + \epsilon \left\langle c_{\kk}^{(0)*} c_{\kk}^{(1)} + c.c. \right\rangle + \epsilon^2 \left( \left\langle \left| c_{\kk}^{(1)} \right|^2 \right\rangle + \left\langle c_{\kk}^{(0)*} c_{\kk}^{(2)} + c.c. \right\rangle  \right) + \textit{O}(\epsilon^3)
\end{equation}
at time $T$, where $\left\langle \cdot \right\rangle$ denotes an ensemble average over possible initial conditions. To this end, we need a closure hypothesis to compute the correlations. In WWT, we use the random phase and amplitude (RPA) hypothesis, which considers that waves have initially random and independent amplitudes and phases, and the phases uniformly distributed in $[0, 2 \pi[$. Within this approximation, we have \citep{nazarenko_wave_2011}
\begin{align}
	\nonumber
	\left\langle c_1 c_2 \right\rangle &= 0, ~~~~~ \left\langle c_1 c_2^* \right\rangle = \left\langle |c_1^{(0)}|^2 \right\rangle ~ \delta_{12}, \\
	\label{eq:RandomPhases}
	\left\langle c_1 c_2 c_3 \right\rangle &= \left\langle c_1 c_2 c_3^* \right\rangle = 0, ~~~~ \left\langle c_1 c_2 c_3 c_4 \right\rangle = \left\langle c_1 c_2 c_3 c_4^* \right\rangle = 0, \\
	\nonumber
	\left\langle c_1 c_2 c_3^* c_4^* \right\rangle &= \delta_{13} \delta_{24} \left\langle |c_1^{(0)}|^2 \right\rangle \left\langle |c_2^{(0)}|^2 \right\rangle + \delta_{14} \delta_{23} \left\langle |c_1^{(0)}|^2 \right\rangle \left\langle |c_2^{(0)}|^2 \right\rangle - \delta_{12} \delta_{13} \delta_{14} \left\langle |c_1^{(0)} |^4\right\rangle.
\end{align}

It is then possible to compute all the terms in (\ref{eq:WaveActionExpansion}) using (\ref{eq:InteractionVariableExpansionStart}-\ref{eq:InteractionVariableExpansionEnd}) and the previous expressions (\ref{eq:RandomPhases}). The second term in the r.h.s in (\ref{eq:WaveActionExpansion}) is
\begin{equation}
	\left\langle c_{\kk}^{(0)*} c_{\kk}^{(1)} + c.c. \right\rangle = 0
\end{equation}
because the triple correlations vanish under the RPA hypothesis. The third term is
\begin{align}
	\left\langle \left| c_{\kk}^{(1)} \right|^2 \right\rangle &= 2 \sum\limits_{1,2} ~ |\Votk|^2 ~ \delta_{12}^{\kk} ~ |\Gamma_T(\omega_{12}^{\kk})|^2 ~ \left\langle |c_1^{(0)}|^2 \right\rangle  \left\langle |c_2^{(0)}|^2 \right\rangle  \\
	\nonumber
	&+ 4 \sum\limits_{1,2} ~ |\Votk|^2 ~ \delta_{\kk2}^1 ~ |\Gamma_T(\omega_{\kk2}^{1})|^2 ~ \left\langle |c_1^{(0)}|^2 \right\rangle  \left\langle |c_2^{(0)}|^2 \right\rangle \\
	\nonumber
	&+ \sum\limits_{1} ~ |V_{11}^{\kk}|^2 ~ \delta_{11}^{\kk} ~ |\Gamma_T(\omega_{11}^{\kk})|^2 ~ \left\langle |c_1^{(0)}|^4 \right\rangle \\
	\nonumber
	&+ 4 \sum\limits_{1,3} ~ V_{11}^{\kk}  V_{33}^{\kk} ~ \delta_{\kk1}^1 \delta_{\kk3}^3 ~ \Gamma_T^*(\omega_{\kk1}^{1}) \Gamma_T(\omega_{\kk3}^{3}) ~ \left\langle |c_1^{(0)}|^2 \right\rangle  \left\langle |c_3^{(0)}|^2 \right\rangle \\
	\nonumber
	&+ 4 \sum\limits_{1} ~ |V_{11}^{\kk}|^2 ~ \delta_{\kk1}^{1} ~ |\Gamma_T(\omega_{\kk1}^{1})|^2 ~ \left\langle |c_1^{(0)}|^4 \right\rangle.
\end{align} 
At this point, we see that the last 3 lines will lead to interactions with shear modes. Generally speaking, all terms where the same index is repeated more than two times in a $\delta$ (e.g. $\delta_{11}^{\kk}$, $\delta_{1\kk}^{1}$, ...) correspond to interactions with shear modes in the limit $\epsilon \rightarrow 0$. To obtain the wave kinetic equation, it is therefore sufficient to consider only the two first terms of the above equation in the following computations. The last term to compute in (\ref{eq:WaveActionExpansion}) is
\begin{align}
	\left\langle c_{\kk}^{(0)*} c_{\kk}^{(2)} + c.c. \right\rangle = &- 8 \sum\limits_{1,2} ~ \delta_{12}^{\kk} ~ \Votk \Vkot ~ \Real \left[ \Lambda_T(-\omega_{21}^{\kk}, \omega_{12}^{\kk}) \right] ~ \left\langle |c_1^{(0)}|^2 \right\rangle  \left\langle |c_{\kk}^{(0)}|^2 \right\rangle \\
	\nonumber
	&- 8 \sum\limits_{1,2} ~ \delta_{\kk2}^1 ~ \Votk \Vkto ~ \Real \left[ \Lambda_T(\omega_{\kk2}^{1}, -\omega_{\kk2}^{1}) \right] ~ \left\langle |c_2^{(0)}|^2 \right\rangle  \left\langle |c_{\kk}^{(0)}|^2 \right\rangle \\
	\nonumber
	&+ 8 \sum\limits_{1,2} ~ \delta_{\kk2}^1 ~ \Votk \Vkot ~ \Real \left[ \Lambda_T(\omega_{\kk2}^{1}, -\omega_{\kk2}^{1}) \right] ~ \left\langle |c_1^{(0)}|^2 \right\rangle  \left\langle |c_{\kk}^{(0)}|^2 \right\rangle
\end{align} 
where we have omitted interactions with the shear modes. The wave action spectrum (\ref{eq:WaveActionSpectrum}), therefore, satisfies the equation
\begin{equation}
	\label{eq:WaveActionDiscreteEquation}
	n_{\kk}(T) - n_{\kk}(0) = \left( \frac{2 \pi}{L} \right)^3 \epsilon^2 \left( \left\langle \left| c_{\kk}^{(1)} \right|^2 \right\rangle + \left\langle c_{\kk}^{(0)*} c_{\kk}^{(2)} + c.c. \right\rangle  \right) + \textit{O}(\epsilon^3).
\end{equation}

The next step is to take the infinite size limit $L \rightarrow \infty$, which is achieved by replacing $\sum\limits_{1,2} \rightarrow \left(\frac{L}{2 \pi}\right)^{6} ~ \int ~  \diff^3\kk_1 ~ \diff^3\kk_2$ and $\delta_{12}^{\kk} \rightarrow \left(\frac{2 \pi}{L}\right)^3 ~ \delta(\kk - \kk_1 - \kk_2 )$ in equation (\ref{eq:WaveActionDiscreteEquation}). Secondly, we take $\epsilon \rightarrow 0$ such that $\lim\limits_{T \rightarrow \infty} \left| \Gamma_T(\omega) \right|^2 = 2 \pi T ~ \delta(\omega)$ and $\lim\limits_{T \rightarrow \infty} \Real \left[ \Lambda_T(-\omega,\omega) \right] = \pi T ~ \delta(\omega)$. In the end, we obtain
\begin{align}
	\frac{n_{\kk}(T) - n_{\kk}(0)}{T} &= 4 \pi \epsilon^2 \int ~  \diff^3\kk_1 ~ \diff^3\kk_2 ~ \\
	\nonumber
	&\left\{|\Votk|^2 ~ \delta(\kk - \kk_1 - \kk_2 ) ~ \delta(\ok - \oo - \ot ) ~ n_1(0) n_2(0) \right. \\
	\nonumber
	&+ 2 ~ |\Votk|^2 ~ \delta(\kk_1 - \kk - \kk_2 ) ~ \delta(\oo - \ok - \ot ) ~ n_1(0) n_2(0) \\
	\nonumber
	&- 2 ~ \Votk \Vkot ~ \delta(\kk - \kk_1 - \kk_2 ) ~ \delta(\ok - \oo - \ot ) ~ n_1(0) n_{\kk}(0) \\
	\nonumber
	&- 2 ~  \Votk \Vkto ~ \delta(\kk_1 - \kk - \kk_2 ) ~ \delta(\oo - \ok - \ot ) ~ n_2(0) n_{\kk}(0) \\
	\nonumber
	&\left. + 2 ~ \Votk \Vkot ~ \delta(\kk_1 - \kk - \kk_2 ) ~ \delta(\oo - \ok - \ot ) ~ n_1(0) n_{\kk}(0) \right\} + \textit{O}(\epsilon^3).
\end{align}

Using the approximation $\frac{n_{\kk}(T) - n_{\kk}(0)}{T} \simeq \dot{n}_{\kk}$ and rearranging the terms allow us to obtain the kinetic equation for weakly nonlinear internal gravity waves (\ref{eq:KineticEquation}-\ref{eq:KineticEquationQ}).

\bibliographystyle{jfm}
\bibliography{biblio}

\end{document}